\documentclass[twocolumn,twoside]{revtex4}
\usepackage{graphicx}
\usepackage{fancyhdr}
\begin{document}

\title{\begin{boldmath}Theory review for hadronic corrections to $g-2$\end{boldmath}}

%

\author{Maarten Golterman}
\affiliation{San Francisco State University, San Francisco, CA 94132, USA\\
and\\ Department of Physics and IFAE-BIST, Universitat Aut\`onoma de Barcelona, 
E-08193 Bellaterra, Barcelona, Spain}

\begin{abstract}
We review Standard-Model evaluations of hadronic contributions to the muon anomalous magnetic moment $g-2$.
Most space is devoted to the hadronic vacuum polarization contribution, in view of the discrepancy between the
data-based dispersive evaluation and the lattice evaluation by the BMW collaboration.
\end{abstract}

\maketitle

\thispagestyle{fancy}


\section{Introduction}
Let us begin with a quick overview of the recent history of the muon anomalous magnetic moment,
$a_\mu=(g-2)/2$, beginning with the high precision measurement of Ref.~\cite{BNL}, the final result of which appeared
in 2006.   Then, in 2020, the many efforts of the theory community resulted in a White Paper (WP)  under
the aegis of the $g-2$ Theory Initiative \cite{WP}, in which the value predicted by the Standard Model (SM) was presented, and 
found to deviate from the BNL experimental value by $3.7\sigma$.   A new experiment at Fermilab confirmed
the BNL result, with a similar precision (based on $6\%$ of the total expected data for the Fermilab experiment),
tightening the discrepancy between the experimental and SM values, now equalling $251(59)\times 10^{-11}$, to $4.2\sigma$.   

The hadronic vacuum polarization (HVP) contribution to $a_\mu$ of Ref.~\cite{WP} is based on a dispersive ``data-based'' approach
combining the world's data for hadronic electroproduction.  At the time  the WP appeared, 
lattice computations of the HVP contribution, while in agreement with the data-based approach, 
were not precise enough to be competitive.   However, meanwhile a new lattice computation of the 
HVP contribution to $a_\mu$ appeared \cite{BMW}, which, if all other contributions to $a_\mu$ are taken to be the values 
obtained in the WP, leads to a SM prediction for $a_\mu$ $2.1\sigma$ larger than the WP value, and only $1.5\sigma$
below the experimental value \footnote{Reference~\cite{BMW} appeared too late for inclusion in the WP value for the SM.}.
Clearly, the theory community's task is to see whether this discrepancy persists, and if so, understand its origin.

In this talk, I review the status of SM evaluations of the hadronic contributions to $a_\mu$.   Section~\ref{HLbL} 
briefly review the status of the hadronic light-by-light (HLbL) contribution, reporting on and updating the estimate presented in the
WP.   I then devote most of this brief review to the HVP contribution, in view of the puzzle described above, focusing on
the recent progress in lattice QCD, in Sec.~\ref{HVP}.   The recent status of the data-based approach is summarized in
Sec.~\ref{DD}, and the lattice approach is discussed, in more detail, in Secs.~\ref{latt} to \ref{other}.  Section~\ref{concl} contains my conclusions.

The goal of this talk is to discuss developments more recent than those
included in the WP, notably the lattice computation of the HVP contribution, and I refer to Ref.~\cite{WP} for an extensive review of both theory and the many other ingredients that together yield the SM prediction for $a_\mu$.

\section{Hadronic light-by-light contribution \label{HLbL}}
\begin{figure}[h]
\centering
\includegraphics[width=80mm]{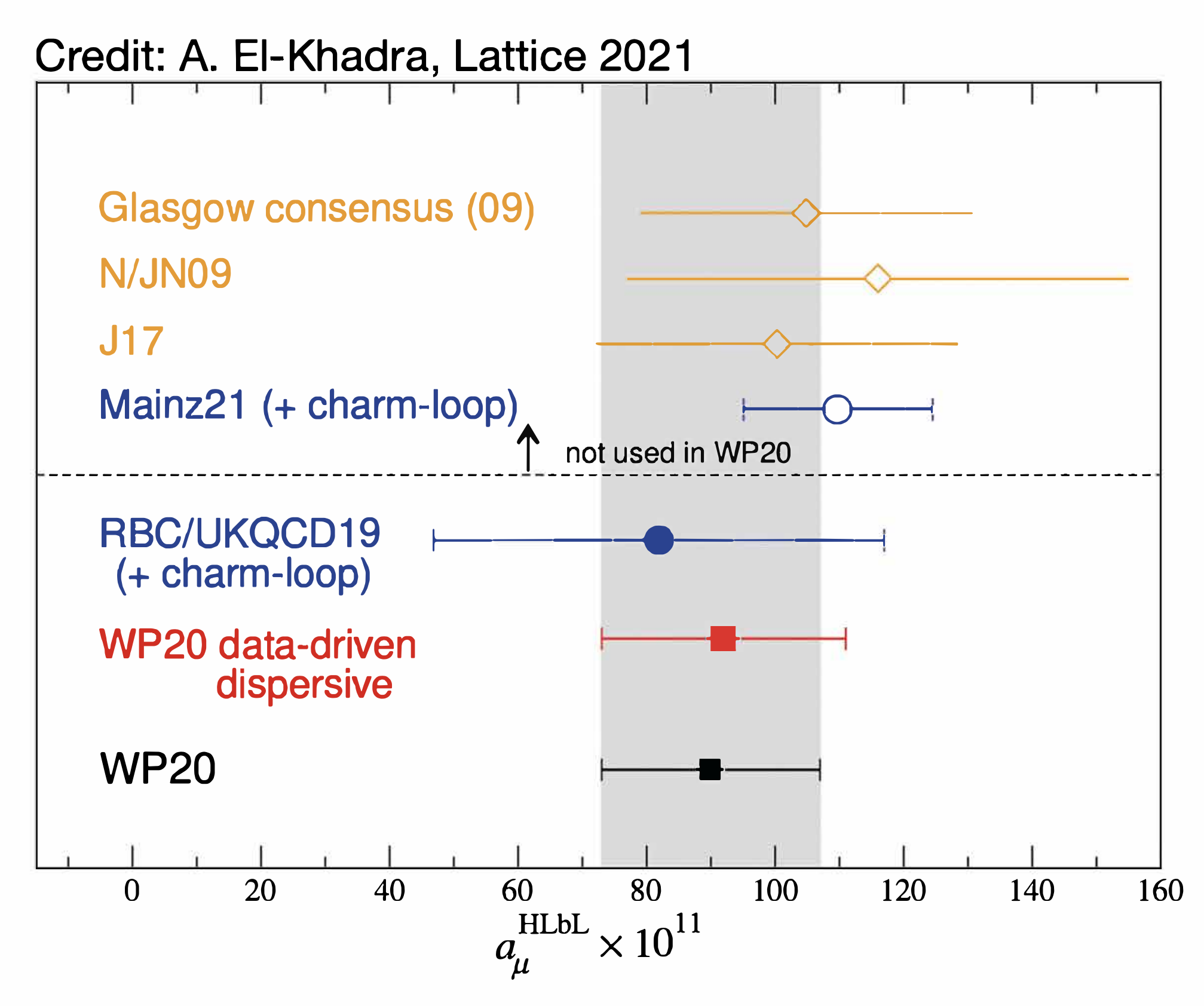}
\caption{Estimates of the HLbL contribution to $a_\mu$ (figure from Ref.~\cite{AEK})} \label{fig:HLbL}
\end{figure}
Figure~\ref{fig:HLbL} shows recent values of the HLbL contribution to $a_\mu$.   The yellow results (open diamonds) above
the horizontal dotted line
show model estimates from Refs.~\cite{Glasgow,JN,J17}.   They are not included in the WP estimate, because the model
approach prevents a reliable estimate of errors.   The WP estimate (black square at the bottom) is the weighted average of
the RBC/UKQCD19 lattice result \cite{RBCHLbL} and a data-based result using dispersive and other phenomenological 
techniques based on Refs.~\cite{J17,HLbLdisp}, 
which are in good agreement with each other.   The lattice result from Mainz (open circle just above the dotted line) \cite{MainzHLbL}
appeared after the WP, but is in good agreement with the other two deteminations on which the WP estimate is based.
Averaging the two results included in the WP estimate with the new Mainz result leads to \footnote{This is my average, not
that of the $g-2$ Theory Initiative.}
\begin{equation}
\label{HLbLres}
a_\mu^{\rm HLbL}=101(11)\times 10^{-11}
\end{equation}
(including the small NLO HLbL contribution),
slightly larger than the WP value $92(18)\times 10^{-11}$, but the small increase does not significantly 
change the discrepancies 
discussed in the Introduction.

The contributions from the $\pi$, $\eta$ and $\eta'$ poles, as well as pion and kaon loops and the contribution from
$S$-wave $\pi\pi$ rescattering have now been evaluated from data using dispersive techniques; these calculations
do not invoke models.   They constitute about 3/4 of the total HLbL contribution.   The remaining contributions
(short distance, axial-vector, etc.) still have to be estimated employing models.   Errors on these latter contributions
are estimated conservatively, and added linearly; they account for most of the total error in the ``data-based
dispersive'' result shown in the figure.  Likewise, improvements are also possible for the lattice determinations,
but again errors on the present determinations are likely to be conservative as well.
At present, there is little reason to expect the central value to move much outside its current error bar when more
complete data-based or lattice results become available.

\section{Hadronic vacuum polarization contribution \label{HVP}}
We next consider the HVP contribution.   The dispersive approach is briefly reviewed in 
Sec.~\ref{DD}, and we then discuss the lattice approach in some more detail in Sec.~\ref{latt} and
subsequent sections.

\subsection{Data-based approach \label{DD}}
The leading-order (LO) HVP contribution to $a_\mu$ can be expressed as \cite{HVPorig}
\begin{eqnarray}
\label{amudisp}
a_\mu^{\rm HVP,LO}&=&\frac{\alpha^2 m_\mu^2}{9\pi^2}\int_{m_\pi^2}^\infty \frac{\hat{K}(s)}{s^2}\,R(s)\,ds\ ,\\
R(s)&=&\frac{\sigma^0(e^+e^-\to\mbox{hadrons}(\gamma)}{4\pi\alpha^2/(3s)}\ ,
\end{eqnarray}
where $\hat{K}(s)$ is a known kernel which increases from about 0.39 at $s=m_\pi^2$ to 1 at $s=\infty$,
and $\sigma^0(e^+e^-\to\mbox{hadrons}(\gamma)$ is the (bare) cross section for electroproduction \footnote{Which
contains higher orders in $\alpha$ because of final-state radiation.}.

Two groups in particular, DHMZ \cite{DHMZ} and KNT \cite{KNT}, have combined the many experimental results
for $e^+e^-\to\mbox{hadrons}(\gamma)$ to obtain $R(s)$.   At low energies ($\sqrt{s}\le 1.8$~GeV for DHMZ and
$\sqrt{s}\le 1.937$~GeV for KNT), $R(s)$ is obtained by combining all measured exclusive channels, while at higher
energies
QCD perturbation theory is used for non-resonant contributions (DHMZ), or a combination of inclusive data and
perturbation theory (KNT).   For the two- and three-pion channels further constraints follow from analyticity
\cite{anal2pi3pi}.   While there are differences between the DHMZ and KNT results, mainly due to a different treatment 
of the well-known
BaBar--KLOE discrepancy in the two-pion channel \cite{WP}, a common value was obtained for the data-based
WP estimate
\begin{equation}
\label{amuHVPdisp}
a_\mu^{\rm HVP,LO}=6931(40)\times 10^{-11}\ .
\end{equation}
A number of comments is in order:
\begin{itemize}
\item The error in Eq.~(\ref{amuHVPdisp}) takes the BaBar-KLOE discrepancy into account; the discrepancy 
accounts for a contribution of about $28\times 10^{-11}$ to the error (in quadrature).  This discrepancy thus cannot
explain the discrepancy between the WP result and the experimental result.  It can also not explain the difference
with the BMW result.
\item No hadronic $\tau$-decay data have been used \cite{ADH}.   The main reason is that, while in an isospin-symmetric
world, $\tau$-decay channels could be converted to the corresponding electroproduction channels using CVC,
there is no reliable way to estimate the corrections from isospin breaking.   In order to compute these isospin
breaking effects, the lattice may be able to help \cite{latttau}.
\item With new data from SND, CMD, BESIII and Belle II, and an analysis of the full BaBar data set, there is a potential
for reducing the error on the data-based $a_\mu^{\rm HVP,LO}$ determination by a factor of about two \cite{snowmass}.   Of course,
it remains to be seen whether new data will resolve the BaBar-KLOE discrepancy, or whether new discrepancies might
arise.
\item New inclusive electroproduction data from BES~III appeared \cite{BESIII} in the energy region 2.2324 to 3.6710 GeV.
These new data appear to be in some tension with QCD perturbation theory.   However, this is unlikely to impact
$a_\mu^{\rm HVP,LO}$.
\end{itemize}

\subsection{Lattice approach \label{latt}}
We now turn to recent lattice results for $a_\mu^{\rm HVP,LO}$ and related quantities, going into some 
more detail than the other topics covered in this
talk because of the tension between the WP and BMW results.   On the lattice, the most commonly used representation
used is the (Euclidean) ``time-momentum'' representation \cite{BM}
\begin{eqnarray}
\label{amulatt}
a_\mu^{\rm HVP,LO}&=&\int_{-\infty}^\infty dt\,w(t)\,C(t)\ ,\\
C(t)&=&\frac{1}{3}\sum_{i=1}^3\sum_{\vec x}\langle j_i^{\rm EM}(\vec x,t)j_i^{\rm EM}(0)\rangle\ ,\nonumber
\end{eqnarray}
where $j_\mu^{\rm EM}=\frac{2}{3}\bar{u}\gamma_\mu u-\frac{1}{3}\bar{d}\gamma_\mu d+\dots$ is the hadronic
EM current, and $w(t)$ is a known weight, related to $\hat{K}(s)$ in Eq.~(\ref{amudisp}).   This representation is
mathematically equivalent to the Euclidean momentum representation of Refs.~\cite{Blum2003}.

We will in particular focus on the light-quark connected
part, which is obtained by choosing only the up/down part of the EM current, and leaving out contractions where the
quark and antiquark at $(\vec x,t)$ are contracted with each other (and thus also the quark and antiquark at 0, with gluons
connecting the two quark loops).   The light-quark connected part
constitutes about 90\% of $a_\mu^{\rm HVP,LO}$, and appears to be the contribution from which the 
tension originates.   Other contributions are the strange, charm and bottom connected parts and the quark-disconnected part, 
which is numerically small, but not negligible, and is much more expensive to compute.   Generally, results for these other
parts are in reasonable agreement \cite{WP,BMW,BGMPdisc} \footnote{New results for the strange, charm and disconnected
parts have also been obtained in Refs.~\cite{ETMC22,Mainz22}.}.
On the lattice, QED and strong isospin breaking effects are taken into account perturbatively, to first order in $\alpha$
and $m_d-m_u$.   For these contributions we refer to Refs.~\cite{WP,BMW,snowmass}.

We need sub-percent precision, and thus very good control over systematic errors.   The most important systematic
errors in the computation of $a_\mu^{\rm HVP,LO}$ are (not in any specific order)
\begin{itemize}
\item  The large-$t$ behavior of $C(t)$, which becomes very noisy for large $t$.   Since $a_\mu^{\rm HVP,LO}$ receives
an important contribution from the large-$t$ part of $C(t)$, control over this region is particularly important for this quantity.
In principle this error is statistical, but if tricks are used it can acquire a systematic component.  
\item  Finite (spatial) volume effects.   A typical volume $L^3$ with $L\sim 5-6$~fm at the physical pion mass leads
to finite-volume effects of about $3-4\%$.  Most collaborations correct for this with NNLO chiral perturbation theory (ChPT)
or with physically inspired models.    Numerical finite-volume studies were carried out in Refs.~\cite{BMW,PACS19}.
\item  Scale setting.   While $a_\mu^{\rm HVP,LO}$ is dimensionless, the scale of the hadronic physics in $C(t)$
needs to be precisely calibrated with the muon mass, which appears in $w(t)$.  A $1\%$ change in the lattice
spacing leads to a $1.8\%$ change in $a_\mu^{\rm HVP,LO}$, implying that precise scale setting is
important \cite{Mainz17}.
\item In this talk, we will focus on the continuum limit.   I will argue that the continuum limit with staggered
fermions, which underlies several recent lattice computations (including Ref.~\cite{BMW}), is not yet safely
under control.   For discussion of the other systematic errors, I refer to Refs.~\cite{WP,snowmass} and references therein. 
\end{itemize}

\begin{figure}[t!]
\centering
\vskip4ex
\includegraphics[width=75mm]{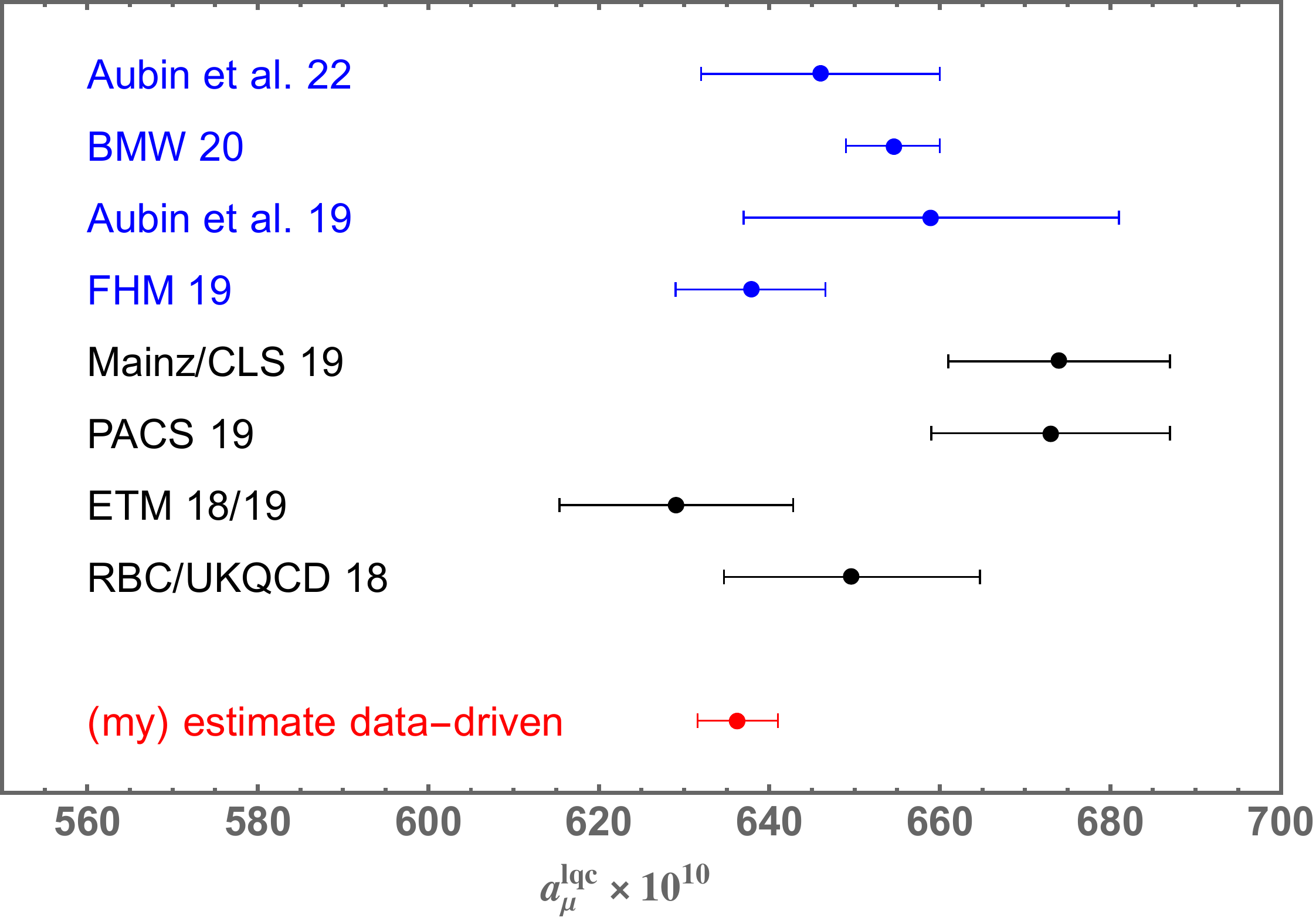}
\caption{Recent determinations of the light quark connected part $a_\mu^{\rm lqc}$ of $a_\mu^{\rm HVP,LO}$,
from Aubin~{\it et al.}~22 \cite{ABGP22}, BMW 20 \cite{BMW}, Aubin~{\it et al.}~19 \cite{ABGP19}, FHM~19 \cite{FHM19}, 
Mainz/CLS~19 \cite{Mainz19}, PACS~19 \cite{PACS19}, ETM~18/19 \cite{ETM19} and RBC/UKQCD~18 \cite{RBC18}.} \label{fig:amulqc}
\end{figure}

Figure~\ref{fig:amulqc} shows recent determinations of the ligh-quark connected contribution.   The blue points
show lattice computations employing staggered fermions, while the red (lowest) point shows my estimate of the
corresponding light quark connected part of the data-based determination \footnote{This was arrived at by
taking the full data-based $a_\mu^{\rm HVP,LO}$ and subtracting the strange quark plus quark-disconnected
part of Ref.~\cite{BGMPdisc}, the charm contribution from Ref.~\cite{WP}, and a combination of QED and SIB
corrections from Refs.~\cite{BMW,KMSIB}.}.

We note that the BMW computation differs from the data-based estimate by about $18\times 10^{-10}$, which
is roughly equal to the difference between the BMW and data-based determinations of the full $a_\mu^{\rm HVP,LO}$,
and that, given the error bars, this difference is significant.   The errors of other lattice computations are still too large
to add more information.

\begin{figure}[t!]
\centering
\vskip4ex
\includegraphics[width=75mm]{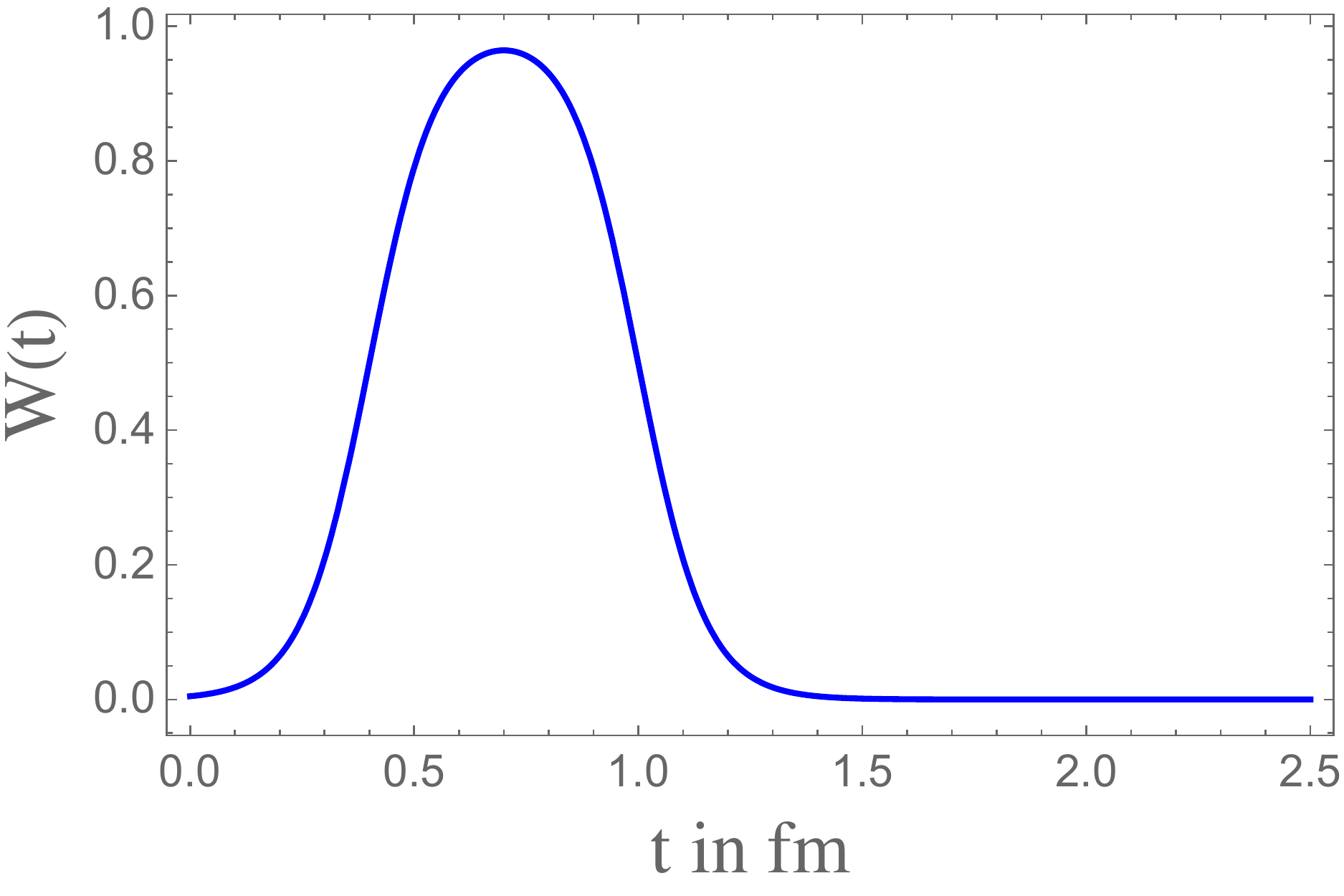}
\caption{The function $W(t)$ used in Eq.~(\ref{amulattwindow}) with $t_0=0.4$~{\rm fm}, $t_1=1.0$~{\rm fm} and
$\Delta=0.15$~{\rm fm}.} \label{fig:window}
\end{figure}
An auxiliary physical quantity that can be computed with much better precision by most lattice collaborations already
at present is the ``intermediate window'' quantity, introduced in Ref.~\cite{RBC18},
\begin{equation}
\label{amulattwindow}
a_\mu^{\rm W}=2\int_0^\infty dt\,W(t)\,w(t)\,C(t)\ ,
\end{equation}
where 
\begin{equation}
\label{W}
W(t)=\frac{1}{2}\left(\tanh\left(\frac{t-t_0}{\Delta}\right)-\tanh\left(\frac{t-t_1}{\Delta}\right)\right)
\end{equation} 
with $t_0=0.4$~fm, $t_1=1.0$~fm and $\Delta=0.15$~fm is the smoothed step-function shown in Fig~\ref{fig:window}.   
The advantage of this quantity is that it
cuts out short- and long-distance effects, which lead to the bulk of the systematic errors in $a_\mu^{\rm HVP,LO}$
on the lattice.   It can be 
computed very precisely, and, importantly, since it is a physical quantity, all lattice computations have to agree.
A caveat is that even though systematic errors are small, they are not negligible.   As this window has support
between 0.4 and 1~fm, it is a relatively short-distance quantity, and thus not accessible to EFT methods such
as chiral perturbation theory to estimate or correct for these systematic errors \cite{ABGP22,ABGPEFT}.

\begin{figure}[t!]
\centering
\vskip4ex
\includegraphics[width=75mm]{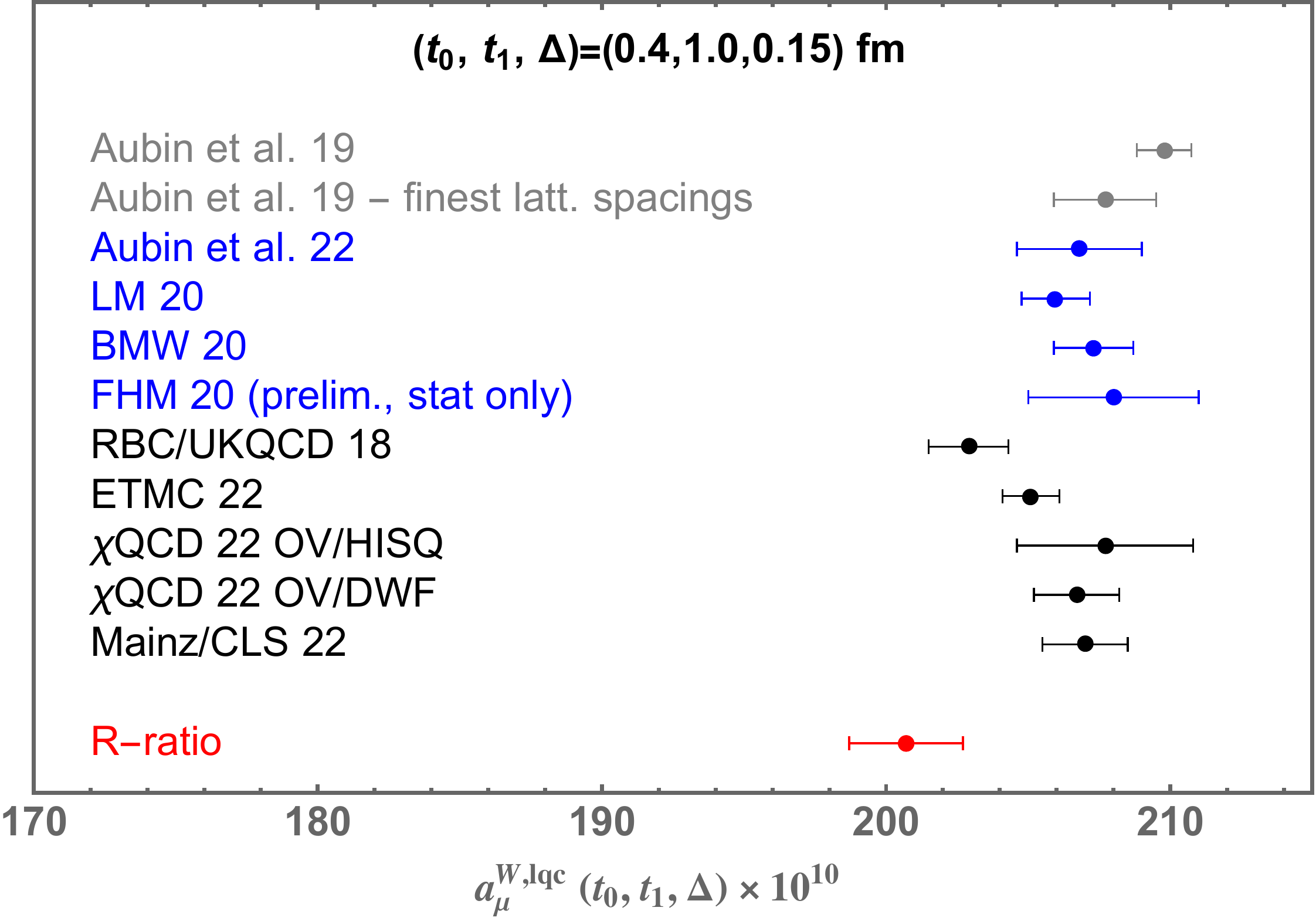}
\caption{Results for the light quark connected part of $a_\mu^{\rm W}$, from
ABGP~{\it et al.}~19 \cite{ABGP19},
ABGP~{\it et al.}~22 \cite{ABGP22}, LM~20 \cite{LM20}, BMW~20 \cite{BMW}, FHM~20 \cite{FHM20},
RBC/UKQCD~18 \cite{RBC18}, ETMC~22 \cite{ETMC22}, $\chi$QCD~22 \cite{chiQCD22} and 
Mainz/CLS~22 \cite{Mainz22}.  The red, lowest data point is an estimate from the data-based approach
provided by C.~Lehner (using data of Ref.~\cite{KNT}).} \label{fig:windowres}
\end{figure}

Figure~\ref{fig:windowres} shows results for the light quark connected part $a_\mu^{\rm W,lqc}$.   
The picture has already changed
considerably since the time of the conference, with the appearance of Refs.~\cite{ETMC22,Mainz22}.  Before,
there appeared to be a tendency for staggered window results to come out higher than other determinations,
such as that of Ref.~\cite{RBC18}.   Now, both new computations yields results which are consistent with the
values of Refs.~\cite{BMW,ABGP22}.  The computations of Ref.~\cite{chiQCD22} had already appeared at the 
time of the conference, but are less conclusive, because they are based on only two lattice spacings, and thus
do not allow for a controlled continuum extrapolation (this also applies to Ref.~\cite{RBC18}).

In short, all computations with three or more lattice spacings now appear to coalesce around a value distinctly
larger than the $R$-ratio based value (red point in Fig.~\ref{fig:windowres}).   Still, this does not mean that we
know the continuum limit (and other systematic errors) to be fully under control.  Below, we will discuss in more detail the continuum limit with 
staggered fermions, comparing the results of Ref.~\cite{BMW} and Ref.~\cite{ABGP22}.    It will be interesting to see what
happens when the continuum limit with staggered fermions is better controlled; in particular, whether the 
agreement among the most recent lattice computations of $a_\mu^{\rm W}$ will hold up.

\subsection{Staggered fermions \label{staggered}}
We begin with a very brief overview of some properties of staggered fermions; for more extensive reviews,
see Refs.~\cite{MGLH,MILC}.  

It is well known that lattice fermions with an exact chiral symmetry exhibit the phenomenon of species 
doubling.   In four dimensions, with hypercubic rotational symmetry, a naively discretized fermion leads
to sixteen degenerate species in the continuum limit.   A staggered fermion is a lattice fermion with only
one (complex) component per site, with one exact chiral symmetry if the (single-site) mass is taken to zero.   The sixteen
one-component doublers organize themselves into four ``tastes'' of Dirac fermions in the continuum limit; on the lattice
the taste and spin components can be thought of being spread out over $2^4$ lattice hypercubes.   The group
of hypercubic rotations mixes spin and taste components, and, on the lattice, only a discrete
subgroup of the continuum $SO(4)_{\rm rot}\times SU(4)_{\rm taste}$ symmetry survives.   The full
$SO(4)_{\rm rot}\times SU(4)_{\rm taste}$ is only restored in the continuum limit.

If we use one staggered fermion for the up quark and one for the down quark, the continuum limit contains
four tastes of up and four tastes of down quarks.   There are thus sixteen charged pions.   Because of the
exact chiral symmetry, one of these pions is a Nambu--Goldstone boson (NGB), but the complete set of pions
breaks up into eight non-degenerate representations of the lattice ``rest-frame'' symmetry group, 
with the exact NGB being the lightest pion.   They 
only become degenerate with the exact NGB in the continuum limit; the
non-degeneracy on the lattice is referred to as ``taste splitting.''

\begin{figure*}[t]
\centering
\includegraphics[width=75mm]{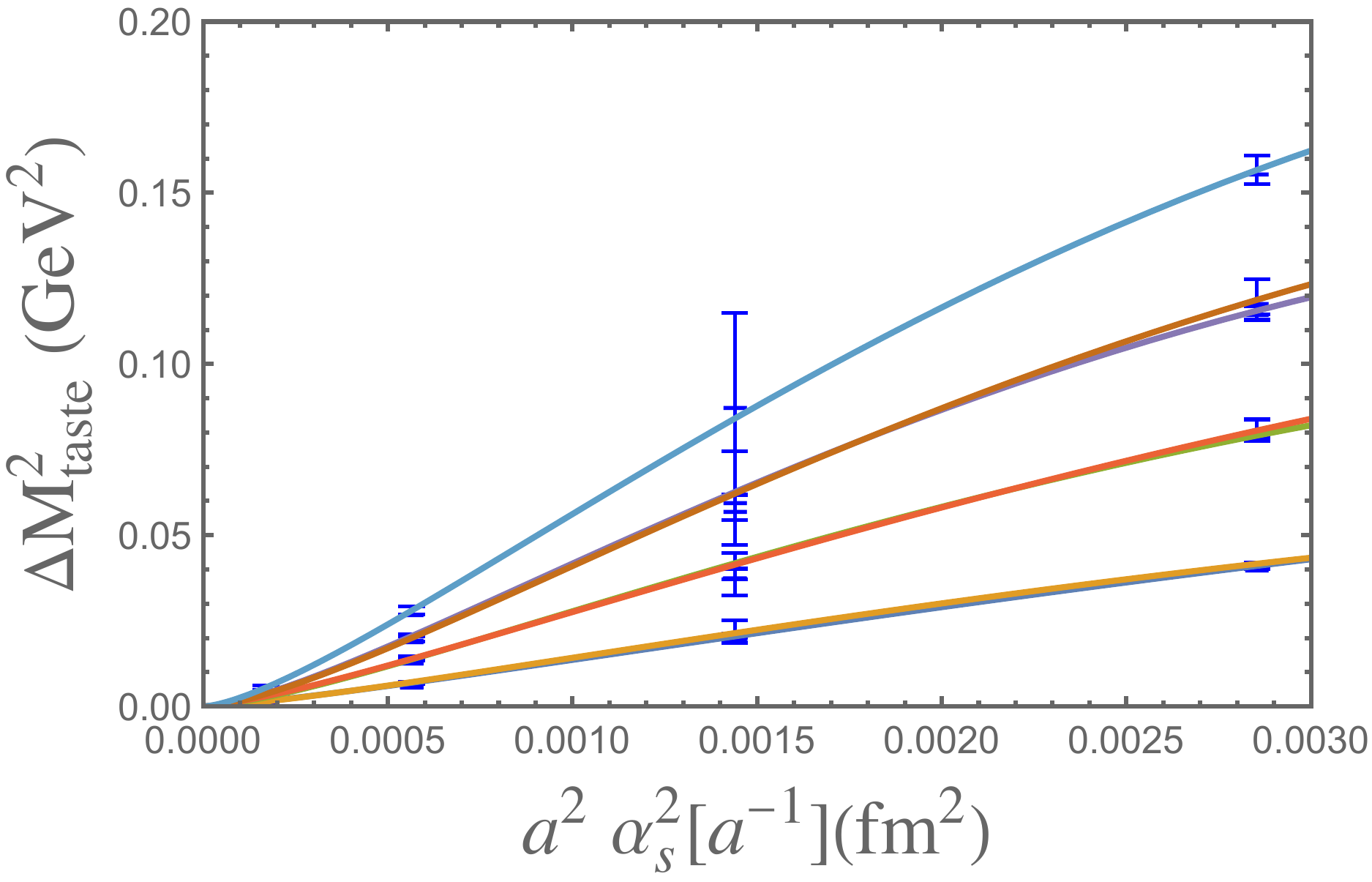}
\includegraphics[width=75mm]{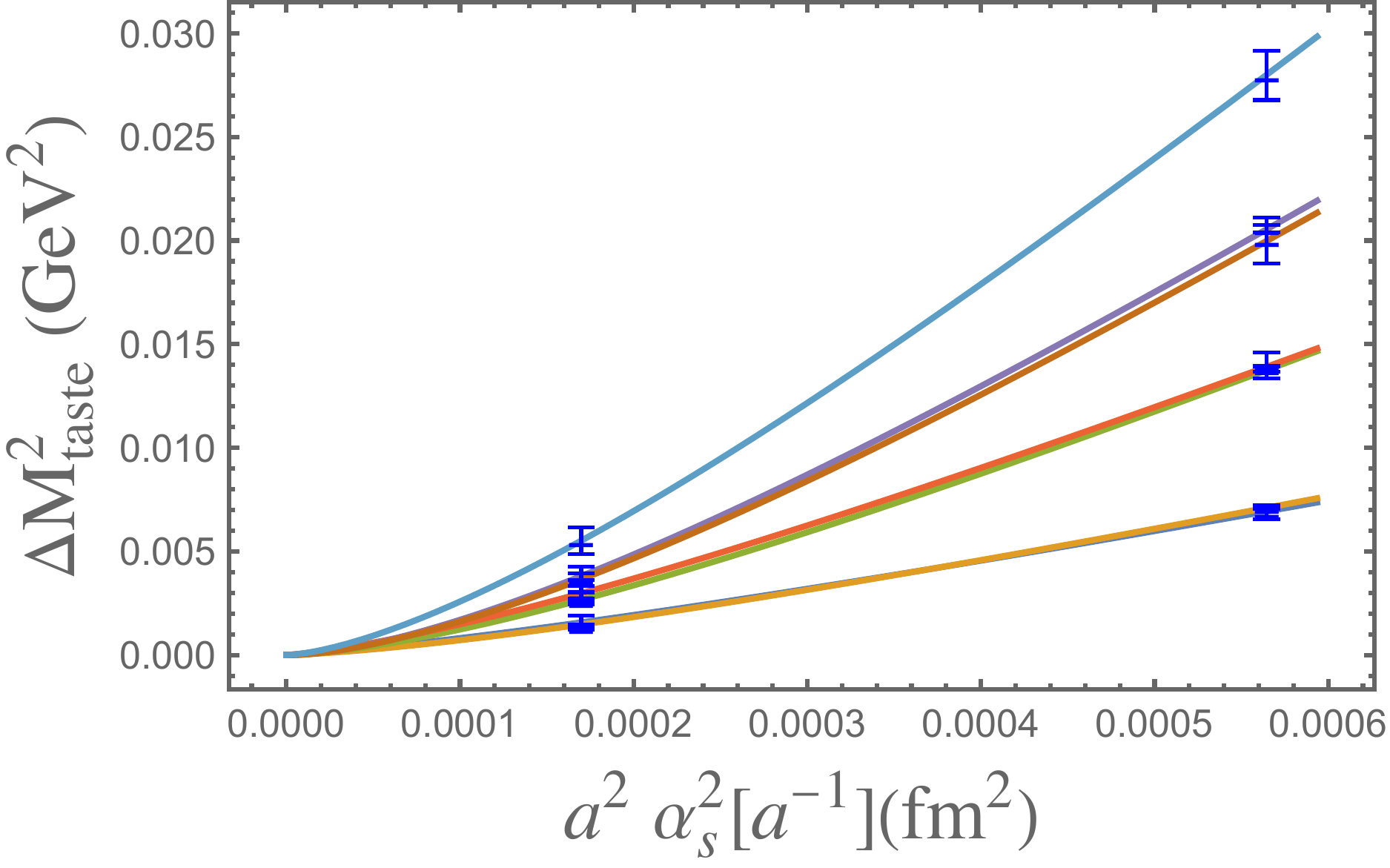}
\caption{Pion taste splittings as a function of $\alpha^2_s a^2$ \cite{ABGP22}.  The right panel zooms in on the lower
left corner of the left panel.   The exact Nambu--Goldstone pion is within a few percent of the physical mass.
Pion mass spectrum courtesy of the MILC collaboration.} \label{fig:taste}
\end{figure*}  

Figure~\ref{fig:taste} shows the pion taste splittings $\Delta M^2_{\rm taste}=M^2_X-M^2_{\rm NGB}$ where $X$ runs over all
non-NGB pions, as a function of $\alpha^2_s(1/a) a^2$ for HISQ \cite{HISQ} staggered fermions
\cite{ABGP22}.   Staggered ChPT (SChPT) \footnote{For a review, see Ref.~\cite{MGLH}.} predicts that taste splittings are
of order $\alpha^2_s a^2$, and that, to that order in the lattice spacing, the $SU(4)$ taste symmetry is only partially broken to
$SO(4)$, while it breaks down completely to the lattice symmetry group at order $a^4$ and beyond \cite{LS}.   In view of this, Fig.~\ref{fig:taste}
presents a puzzle:  the (five) approximate $SO(4)$ multiplets predicted by leading-order SChPT are clearly seen, but the taste
splittings are very non-linear in $\alpha^2_s a^2$.   Order $a^4$ and $a^6$ terms are needed to obtain good fits, and in fact the
coefficient of the term of order $\alpha^2_s a^2$ is consistent with zero.   The upshot is that taste splittings show large scaling
violations, and these scaling violations are very non-linear in $a^2$.   This strongly suggests that other
quantities, such as $a_\mu^{\rm W}$, which are very sensitive to taste splittings, are, at the lattice
spacings shown in Fig.~\ref{fig:taste}, not in the regime of linear behavior in $\alpha^2_s a^2$, making the continuum extrapolation
more difficult.   Clearly, a more detailed study of taste splittings for the various staggered actions in use is called for.

\begin{figure*}[t]
\centering
\includegraphics[width=75mm]{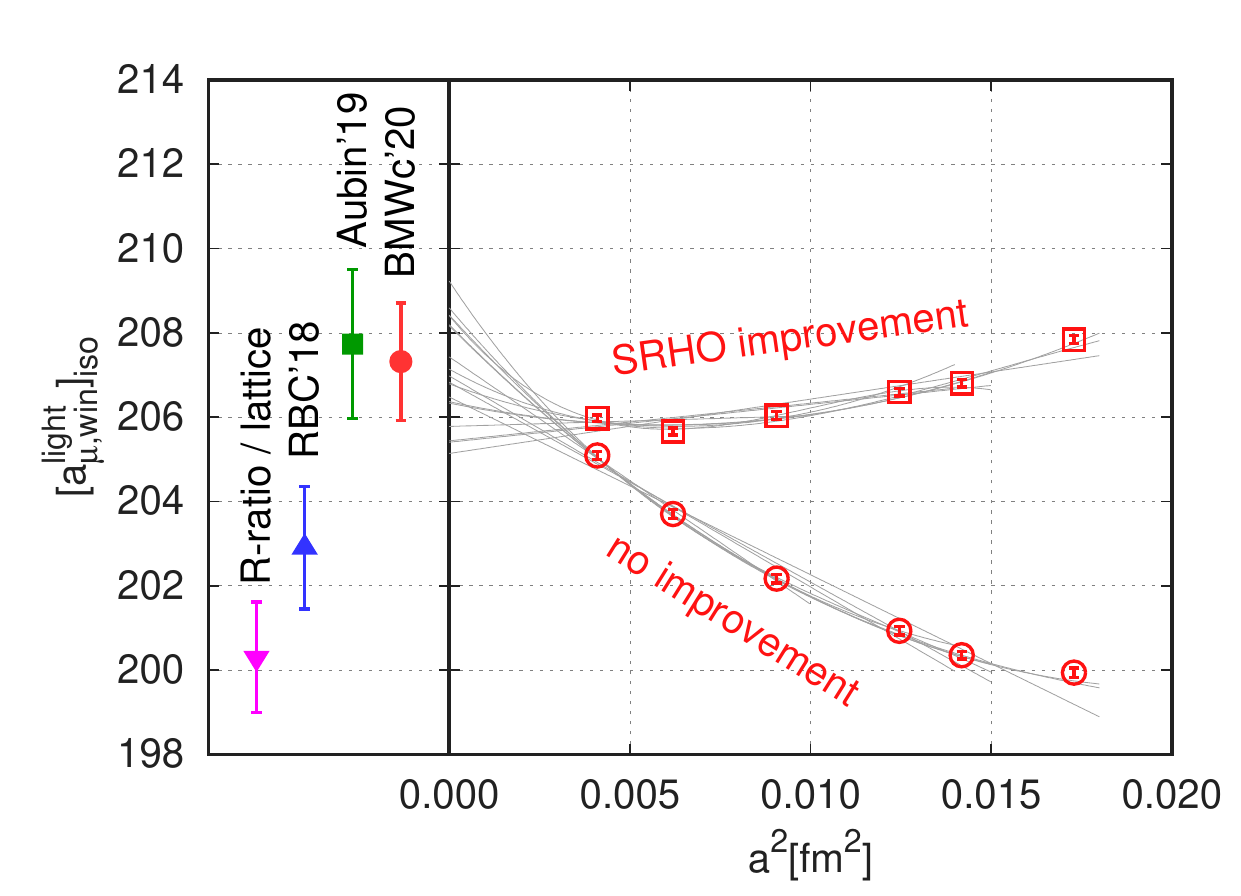}
\hskip1cm
\includegraphics[width=75mm]{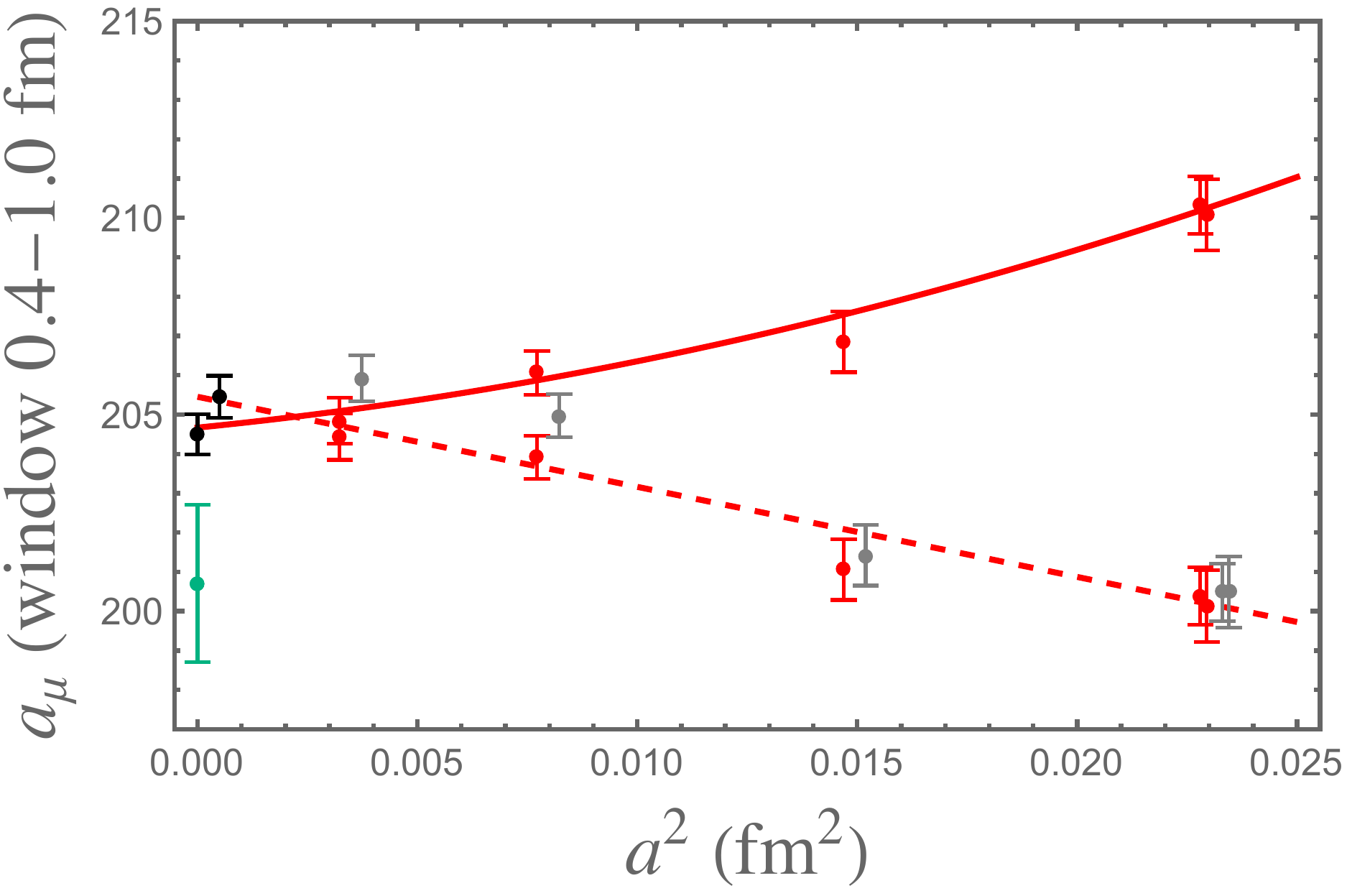}
\caption{Results for $a_\mu^{W,lqc}$ from BMW~20 \cite{BMW} (left panel) and 
ABGP~22 \cite{ABGP22} (right panel).  The latter used HISQ configurations generated by the MILC collaboration
\cite{MILCHISQ}. In the right panel, the solid curve is with SRHO
improvement, the dashed curve without.   The $R$-ratio based estimate is shown as a magenta
inverted triangle in the left panel, and as the green lower point at $a^2=0$ in the right panel.} \label{fig:amuW}
\end{figure*}  

Figure~\ref{fig:amuW} shows results for $a_\mu^{\rm W,lqc}$ from Ref.~\cite{BMW} (left plot) and from
Ref.~\cite{ABGP22} (right plot).   Both figures show only some of the fits considered in Refs.~\cite{BMW,ABGP22},
and we refer to the original papers for a complete discussion.
The lower curve in each plot shows a continuum extrapolations for the
unimproved data, while the upper curve shows the same data, corrected for the pion taste splittings using a
physical model (the ``SRHO'' model, first proposed in Ref.~\cite{HPQCD16}).   (All red data points are
corrected for finite-volume and pion-mass mistuning effects.)  These two continuum 
extrapolations should agree with each other, since the difference is an order-$a^2$ effect.   It seems clear
from both figures that this agreement is not perfect, especially if one takes into account that the two
extrapolations are highly correlated.   (The difference is taken into account in the systematic error.)

There are several observations to make about these results.   First, 
as already shown in Fig.~\ref{fig:windowres}, both computations lead
to consistent values for $a_\mu^{\rm W,lqc}$, and they are in tension with the $R$-ratio based estimate
\footnote{There is no space here to review the detailed systematic error analysis for each 
computation, for which we refer to the original papers.}.
Then, the improved data points, in which
the lattice-spacing artifacts due to taste breaking are corrected for, are not linear in $a^2$, in both cases.
Furthermore, taste-breaking effects are large, and significant even for the smallest $a^2$ value in each
plot.   In fact, even though the computations of Ref.~\cite{BMW} and Ref.~\cite{ABGP22} are based on
different (improved) staggered actions, it is striking how similar both figures are.   Taste-breaking
effects, as measured by the difference of the two curves in each plot, are approximately the same 
size for the two computations at similar values of $a^2$.

\begin{figure}[t!]
\centering
\vskip4ex
\includegraphics[width=75mm]{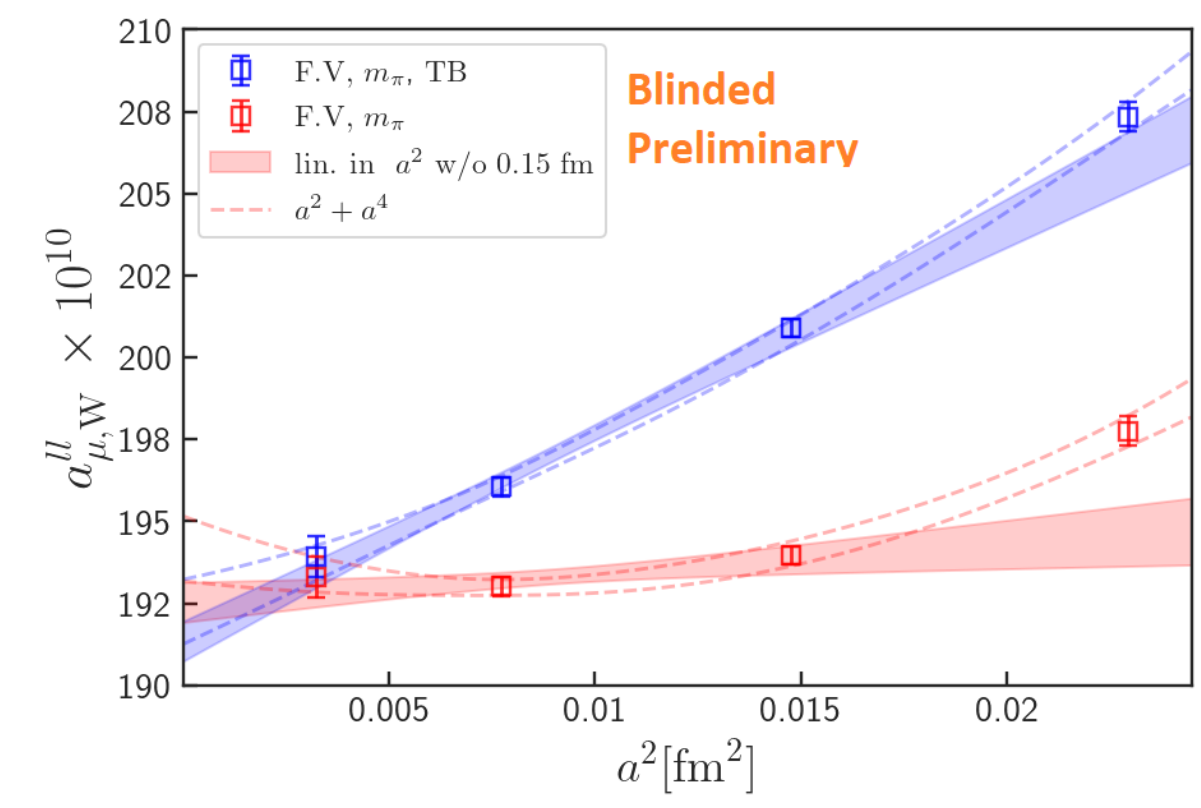}
\caption{Blinded preliminary window result of FHM~20 \cite{FHM20}.   Units on the vertical 
axis are arbitrary.  The colored bands show linear fits omitting the $a=0.15$~{\rm fm} points
from the fits.} \label{fig:FHM}
\end{figure}

Figure~\ref{fig:FHM} shows a still blinded plot of the window results of Ref.~\cite{FHM20},
also computed with the HISQ ensembles of Ref.~\cite{MILCHISQ}.
Again, the lower (red) curve is without correcting for taste breaking, while the upper (blue) curve
is corrected.   Remarkably, now the slope of the curve with the taste-breaking correction is the
steeper one, unlike what is seen in Fig.~\ref{fig:amuW}.   The  most likely explanation is that
Refs~\cite{BMW,ABGP22} used a conserved current, while Ref.~\cite{FHM20} used a local
current (with the appropriate finite renormalization).   Both sets of data points show 
non-linear behavior with $a^2$, and this appears to remain true when the $a=0.15$~fm data points are
removed.

From these results, and from the discussion of Fig.~\ref{fig:taste}, we conclude that staggered 
computations of $a_\mu^{\rm W}$, and, by extension, $a_\mu^{\rm HVP,LO}$, should be carried out
at smaller lattice spacings than the current smallest available lattice spacing $a\sim 0.06$~fm.
Adding smaller lattice spacings than the currently smallest lattice spacing of about 0.06~fm
would allow dropping the data points with lattice spacings larger than 0.1~fm, which are clearly 
in the regime of non-linear behavior in $a^2$.

\subsection{Other fermion methods \label{other}}
Other fermion discretizations have been used for the computation of $a_\mu^{\rm HVP,LO}$
and $a_\mu^{\rm W}$, notably domain-wall fermions \cite{RBC18}, overlap fermions \cite{chiQCD22},
twisted-mass fermions \cite{ETMC22}, and improved Wilson fermions \cite{Mainz22}.   

Only Ref.~\cite{RBC18}
finds a lower value for $a_\mu^{\rm W}$, consistent with the $R$-ratio based value.   Their result is based
on a linear extrapolation from two lattice spacings, with a rough estimate for the size of possible $a^4$ effects.
New results with a third, smaller, lattice spacing are expected in the near future
with a very small statistical error.   It will be interesting to see where the new RBC/UKQCD value will be
located in an updated version of Fig.~\ref{fig:windowres}.

The new results of Refs.~\cite{ETMC22,Mainz22} are based on three or more lattice spacings, and
confirm the results found in Refs.~\cite{BMW,ABGP19,ABGP22}.   Nevertheless, I believe that the
staggered results should be corroborated (or modified) with computations at smaller lattice spacings,
because of the uncertainties associated with the staggered continuum limit discussed above.  I note
that the smallest lattice spacing of Ref.~\cite{ETMC22} is also close to 0.06~fm.   The smallest 
lattice spacing of Ref.~\cite{Mainz22} is 0.039~fm, but in this case the continuum extrapolation is
combined with phenomenological extrapolations from heavier-than-physical pion masses.  While the analysis of
Ref.~\cite{Mainz22} includes two near-physical pion-mass ensembles with lattice spacings of
about 0.06 and 0.08~fm, their results are based on
ensembles with pion masses up to 422~MeV.   

\section{Conclusion \label{concl}}
This brings me to my conclusions, which, of course, apply to the present situation.  The current
state of the field can 
be expected to evolve rapidly in the near future.
\begin{itemize}
\item  First, I would like to emphasize the importance of the $g-2$ Theory Initiative, which is
a community-based effort to collect and review computations of SM contributions to $a_\mu$,
with the goal of providing and updating a SM value with controlled errors \cite{WP,snowmass}.
\item With regard to hadronic contributions, HLbL appears to be in very good shape, with
agreement between competitive lattice and data-based approaches.   With regard to the 
HVP contribution, there is a tension between the data-based approach summarized in
\cite{WP}, and the lattice computation by the BMW collaboration \cite{BMW}, which is the only
lattice result with a sub-percent error at present.   However,
many collaborations are now computing the ``intermediate window'' quantity ({\it cf.} Fig.~\ref{fig:windowres}).
At present, there appears to be a growing tension between most lattice results for $a_\mu^{\rm W}$
and the data-based value, amounting to roughly half of the discrepancy between the values for
$a_\mu^{\rm HVP,LO}$ obtained by the data-based approach and Ref.~\cite{BMW}.  
\item Nevertheless, it appears that more work is needed to be certain that the continuum limit
is under control.   Certainly for staggered-fermion based computations, which constitute roughly half
of the lattice effort, lattice spacings smaller
than 0.06~fm will be indispensible to extrapolate more reliably to the continuum limit.   As the 
domain-wall fermion based value of Ref.~\cite{RBC18} is now in some tension with most other
lattice results (see Fig.~\ref{fig:windowres}), it is 
important to see whether a third lattice spacing will confirm their earlier result.
\item Finally, while the ``intermediate window'' is a nice quantity from the lattice point of view,
the scales it picks out, between 0.4 and 1.0~fm, are relatively short-distance.   This prevents the
use of EFT methods to correct for systematic effects (such as finite-volume corrections, and 
pion-mass corrections).   While these effects are much smaller for $a_\mu^{\rm W}$ than for
$a_\mu^{\rm HVP,LO}$, they are not negligible.   At least at present, models are thus needed
for these corrections, introducing a -- possibly mild -- model dependence in the results.
In this regard, it may be interesting to consider windows with longer distance scales, which
are accessible to ChPT, as proposed in Ref.~\cite{ABGP22}.
\end{itemize}

\begin{acknowledgments}
First, I would like to thank Jake Bennett and the other organizers for a very nice conference.
I would like to also thank my collaborators and other colleagues in this field for many discussions and
inputs.   I would like to mention Christopher Aubin, Tom Blum, Diogo Boito, Mattia Bruno, Gilberto Colangelo, Aida El-Khadra, 
Gregorio Herdoiza, Martin Hoferichter, Jamie Hudspith, Alex Keshavarzi, Christoph Lehner, Kim Maltman, Aaron Meyer, and, in
particular,  Santi Peris. 
MG is supported by the U.S.\ Department 
of Energy, Office of Science, Office of High Energy Physics, under 
Award DE-SC0013682.
\end{acknowledgments}

\bigskip 

\end{document}